\documentclass[twocolumn,pre,superscriptaddress]{revtex4}
\usepackage{graphicx}
\usepackage{amsmath,amssymb}

\begin{document}

\title{Anomalous properties of heat diffusion in living tissue caused by
branching artery network. Qualitative description}
\author{I. A. Lubashevsky}
\affiliation{Theory Department, General Physics Institute, Russian Academy of
Sciences}
\affiliation{Institute for Applied Problems in Mechanics and
Mathematics, National Academy of Sciences of Ukraine}
\author{V. V. Gafiychuk}
\author{B. Y. Datsko}
\affiliation{Institute for Applied Problems in Mechanics and Mathematics,
National Academy of Sciences of Ukraine}

\date{\today}

\begin{abstract}
We analyze the effect of blood flow through large arteries of peripheral
circulation on heat transfer in living tissue. Blood flow in such arteries
gives rise to fast heat propagation over large scales, which is described in
terms of heat superdiffusion. The corresponding bioheat heat equation is
derived. In particular, we show that under local strong heating of a small
tissue domain  the temperature distribution inside the surrounding tissue is
affected substantially by heat superdiffusion.
\end{abstract}

\maketitle

\section{Introduction. Living tissue as a heterogeneous medium}

Blood flowing through vessels forms paths of fast heat transport in living
tissue and under typical conditions it is blood flow that governs heat
propagation on scales about or greater than one centimeter (for an
introduction to this problem see, e.g., \cite{CH80,SE70}). Blood vessels make
up a complex network being practically a fractal. The larger is a vessel, the
faster is the blood motion in it and, so, the stronger is the effect of blood
flow in the given vessel on heat transfer. Blood flow in capillaries
practically does not affect heat propagation whereas blood inside large
vessels moves so fast that its heat interaction with the surrounding cellular
tissue is ignorable \cite{CH80}. Thus, there should be vessels of a certain
length $\ell _{v}$ that are the smallest ones among the vessels wherein blood
flow affects heat transfer remarkably. The value of $\ell _{v}$ can be
estimated as \cite{BOOK} (see also \cite{CH80,SE70}):
\begin{equation}
\ell _{v}\sim \sqrt{\frac{D}{jfL_{n}}}\,,  \label{ThC:e1.4}
\end{equation}
where $D=\kappa /(c\rho )$ is the temperature diffusivity of the cellular
tissue determined by its thermal conductivity $\kappa $, specific heat $c$,
and density $\rho $, the value $j$ is the blood perfusion rate (the volume of
blood going through tissue region of unit volume per unit time), and the
factor $L_{n}\sim\ln \left( l/a\right) $ is logarithm of the mean ratio of the
individual length to radius of blood vessels forming peripheral circulation.
For the vascular networks made up of the paired artery and vein trees where
all the vessels are grouped into the pairs of the closely-spaced arteries and
veins with opposite blood currents the coefficient $f\sim L_{n}^{-1/2}$
accounts for the counter-current effect
\footnote{%
Initially the factor $f$ was phenomenologically introduced in the bioheat
equation to take into account a certain renormalization of the blood perfusion
rate caused by the counter-current effect \cite{C80,W87a,W87b}. Its
theoretical estimate was later obtained independently by Weinbaum \textit{et
al.} \cite{WXZE97} and Gafiychuk~\&~Lubashevsky \cite{BOOK} (announced for the
first time in \cite{we1}).}. For the vascular networks where the artery and
vein trees are arranged independently of each other the factor $f$ should be
set equal to unity, $f=1$. In particular, for the typical values of the ratio
$l/a\sim 20$--40 \cite{Mch89}, the thermal conductivity $\kappa \sim 7\cdot
10^{-3}$ \thinspace W/cm$\cdot $K, the heat capacity $c\sim 3.5$\thinspace
J/g$\cdot $ K, and the density $\rho \sim 1$ \thinspace g/cm$^{3}$ of the
tissue as well as setting the blood perfusion rate $j\sim 0.3$\thinspace
min$^{-1}$ from (\ref{ThC:e1.4}) we get the estimates $\ell _{v}\sim 4$~mm and
$L_{n}\approx 3-4\,$.

In the mean-field approximation the effect of blood flow on heat transfer is
reduced to the renormalization of the temperature diffusivity, $D\rightarrow
D_{\text{eff}}$, \cite{WJ85} and the appearance of the effective heat sink
$fj$ \cite{CH80,BOOK,WXZE97} in the bioheat equation:
\begin{equation}
\frac{\partial T}{\partial t}=\nabla \left( D_{\text{eff}}\nabla T\right)
-fj(T-T_{a})+q_{T}\,.  \label{eq:BH}
\end{equation}
Here $T(\mathbf{r},t)$ is the tissue temperature field averaged over scales
about $\ell_v$, the parameter $T_{a}$ is the blood temperature insider the
systemic circulation arteries, and the summand $q_{T}(\mathbf{r},t)$ called
below the temperature generation rate is specified by the heat generation rate
$q$ as $q_{T}=q/(c\rho )$. The renormalization of the temperature diffusivity
is mainly determined by the blood vessels of lengths about $\ell_v$ and due to
the fractal structure of vascular networks the renormalization coefficient
$F=D_{\text{eff}}/D$ is practically a constant of unity order, $F\gtrsim 1$
\cite{BOOK}.

\begin{figure}[tbp]
\begin{center}
\includegraphics[width=60mm]{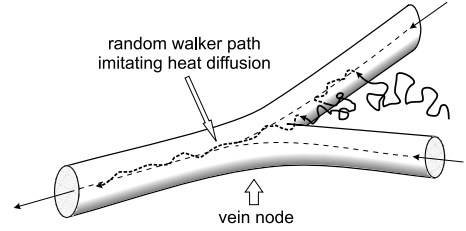}
\end{center}
\caption{Schematic illustration of the effect of blood flow through the vein
tree on heat diffusion imitated by random walks. The figure shows trapping of
a random walker because of getting the internal points of a large vein after
passing the vein node.} \label{Fig1}
\end{figure}

Let us imitate the temperature evolution in terms of random walks whose
concentration is $(T-T_{a})$. Then the part of the vein tree made up of vessels
whose lengths exceed or are about the scale $\ell _{v}$ forms the system of
traps. In fact, blood streams going through the vein tree merge into greater
streams at the nodes (Fig.~\ref{Fig1}). Therefore an effective random walker
after reaching the boundary of one of these veins inevitably will be moved by
blood flow into the internal points of large veins. Then, due to relatively
fast blood motion inside these veins it will be carried away from the tissue
region under consideration, which may be described in terms of the walker
trapping or, what is the same, the heat sink \cite{BOOK}. Since the mean
distance between these veins is determined mainly by the shortest ones, i.e.
by the veins of length $\ell_{v}$ the mean time during which a walker wanders
inside the cellular tissue before being trapped is \cite{BOOK}
\begin{equation}
\tau \sim \frac{\ell _{v}^{2}}{D}L_{n}\,.  \label{eq:Tau}
\end{equation}
In obtaining the given expression we have assumed the vascular network to be
embedded uniformly in the cellular tissue, so the tissue volume $\ell_v^{3}$
falls per one vein (and artery, respectively) of length $\ell_{v}$. Whence it
follows, in particular, that the rate at which the walkers are being trapped by
these vein, i.e. the rate of their disappearance is estimated as $1/\tau $,
leading together with expression~(\ref{ThC:e1.4}) to the heat sink of
intensity $fj$ in the bioheat equation~(\ref{eq:BH}). The characteristic
spatial scale of walker diffusion in the cellular tissue before being trapped
is $\ell _{T}\sim \sqrt{D\tau }$, i.e.
\begin{equation}
\ell _{T}\sim \ell _{v}\sqrt{L_{n}}\sim \sqrt{\frac{D}{jf}}\sim 1\,\text{cm}.
\label{ThC:e1.7a}
\end{equation}
The scale $\ell _{T}$ gives us also the mean penetration depth of heat
penetration into the cellular tissue from a point source or, what is the same,
the widening of the temperature distribution caused by heat diffusion in the
cellular tissue. It is the result obtained within the mean-field approximation.

Beyond the scope of the mean-field theory we meet several phenomena. One of
them is the temperature nonuniformities caused by the vessel discreteness
\cite{Bi86} which can be described assuming the heat sink in
equation~(\ref{eq:BH}) to contain a random component \cite{BOOK}. Another is
fast heat transport over scales exceeding substantially the length $\ell _{T}$
caused by the effect of blood flow through the artery tree. The latter
phenomena is the main subject of the present paper.

\section{Fast heat transport with blood flow through large artery tree}

\begin{figure}[tbp]
\begin{center}
\includegraphics[width=80mm]{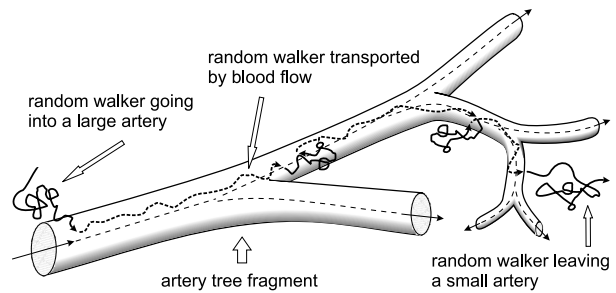}
\end{center}
\caption{Schematic illustration of the anomalously fast heat propagation
caused by blood flow through large artery tree in terms of random walks. The
figure shows the effective walker motion with blood from a large artery to a
small one where the walker leaves it wandering in the cellular tissue.}
\label{Fig2}
\end{figure}

Let at a certain time a random walker wandering in the cellular tissue get a
boundary of a large artery, i.e. an artery of length exceeding $\ell _{v}$. It
should be noted that such an event is of low probability and cannot be
considered within the standard mean-field approximation because the relative
number of large arteries is small. Due to the direction of the blood motion
from larger arteries to smaller ones as well as the high blood flow rate in the
large arteries the walker will be transported fast to one of the arteries of
length $\ell _{v}$ (Fig.~\ref{Fig2}). The blood flow rate in small vessels is
not high enough to affect the walker motion essentially and it has inevitably
to leave this artery and wander in the cellular tissue until being trapped by
the veins of length $\ell _{v}$. Thereby a certain not too large number of
random walkers generated, for example, inside a cellular tissue neighborhood
of a point $\mathbf{r}$ can be found during the time $\tau$ inside a cellular
tissue neighborhood of a point $\mathbf{r}^{\prime }$ at a distance much larger
than $\ell _{v}$, i.e. $\left| \mathbf{r}-\mathbf{r}^{\prime }\right| \gg \ell
_{v}$. The given effect may be regarded as anomalously fast heat diffusion in
living tissue, i.e. heat superdiffusion.

Dealing with heat transfer in living tissue we may confine our consideration
to the peripheral vascular networks typically embedded uniformly into the
cellular tissue, at least at the first approximation \cite{M77}. The latter
statement means, in particular, the fact that for a fixed peripheral vascular
network the vessel collection comprising all the arteries of length $l$ meets
the condition of the volume $l^{3}$ approximately falling per each one of
these arteries. Therefore as is seen in Fig.~\ref{Fig2} a walker going into a
large artery of length $l$ at initial time during the time $\tau $ before
being trapped by the veins can be found equiprobably at each point of the
given artery neighborhood of size $l$. In other words, this walker makes a
large jump of length $l$ that exceeds substantially the mean-field diffusion
length $\ell _{T}$.

In what follows we will analyze the temperature distribution averaged over all
the possible realizations of the vascular network embedding. This enables us to
regard a walker entering a large artery of length $l$ as a random event whose
probability is independent of the walker initial position. Thereby, the
probability $P_{l}$ for a walker to make a large jump over the distance $l$ is
also independent of the spatial coordinates $\mathbf{r}$. It should be noted
that for a fixed realization of the vascular network embedding the probability
$P_{l}$ depends essentially on the spatial coordinates $\mathbf{r}$ and the
heat transfer in living tissue on large scales has to exhibit substantial
dependence on the specific position of the heat sources.

Now we estimate the value of $P_{l}$ assuming the heat sources to be localized
inside a domain $Q_{\mathcal{L}}$ of size $\mathcal{L}$. Two different factors
determine the value of $P_{l}$. First, it is the process of walker trapping by
an artery of length $l>$ $\mathcal{L}$ going through the domain
$Q_{\mathcal{L}}$. If $l<$ $\mathcal{L}$ blood flow in this artery has
practically no effect on heat diffusion. On the average a random walker during
the time $\tau$ travels the distance $\ell _{T}$ in the cellular tissue until
being trapped by the veins of length $\ell _{v}$. So for a walker to enter this
artery and, thus, to leave the domain $Q_{\mathcal{L}}$ with blood flow in it
the walker, on one hand, should be located at initial time inside a cylindrical
neighborhood $Q_{l}$ of the given artery whose radius is about $\ell _{T}$ and
the volume is $\mathcal{L}\ell _{T}^{2}$. On the other hand, it has to avoid
being trapped by the veins of length $\ell _{v}$. The probability of the latter
event is about $(\ell _{v}/\ell _{T})^{2}$. Indeed, a vein of length $\ell
_{v}$ may be treated as a trap of cylindrical form. Thereby in qualitative
analysis the walker trapping can be described in terms of two-dimensional
random walks in the plane perpendicular to the artery under consideration
where the trapping veins are represented by small circular regions. Their
density is about $1/\ell _{v}^{2}$ which directly leads to the latter
estimate. Therefore the total number of walkers leaving the domain
$Q_{\mathcal{L}}$ with blood flow through the given artery per unit time is
\begin{equation}
\frac{1}{\tau }\cdot \left( \frac{\ell _{v}}{\ell _{T}}\right) ^{2}\cdot
\mathcal{L}\ell _{T}^{2}\cdot (T-T_{a})=\frac{D}{L_{n}}\mathcal{L}(T-T_{a})\,.
\label{add1}
\end{equation}
In obtaining~(\ref{add1}) we have taken into account
expression~(\ref{eq:Tau}). Since the trapped walkers spread uniformly over a
region of size $l$ the resulting density of the walker transition rate to a
point $\mathbf{r}$ spaced at a distance about $l$ from the domain
$Q_{\mathcal{L}}$ is
\begin{equation}
g_{l}(\mathbf{r})\sim \frac{D}{L_{n}}\frac{\mathcal{L}}{l^{3}}(T-T_{a})\,.
\label{add2}
\end{equation}
It should be noted that the transition rate $g_{l}(\mathbf{r})$, as it must,
does not depend on the local value of blood perfusion rate.

\begin{figure}[tbp]
\begin{center}
\includegraphics[width=80mm]{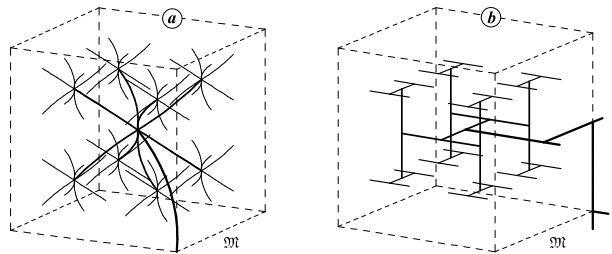}
\end{center}
\caption{Models of the peripheral artery network embedding into the cellular
tissue, ($a$) four-fold node model used in the present analysis and ($b$) a
more realistic dichotomic artery tree uniformly embedded into a cellular
tissue domain $\mathcal{M}$. In the qualitative description of heat transfer
both the models lead to the same result.\label{Fig3}}
\end{figure}

At the second step we should average the obtained transition rate
$g_{l}(\mathbf{r})$ over the possible realizations of the vascular network
embedding. Let us adopt a simplified model for the vascular network shown in
Fig.~\ref{Fig3}$a$ where the vessel lengths $l_{n}$ and $l_{n+1}$ of the
neighboring hierarchy levels $n$ and $n+1$ are related as $l_{n}=2l_{n+1}$.
Figure~\ref{Fig3}$b$ demonstrates a more adequate model for the peripheral
artery tree which, however, within the framework of the present qualitative
analysis may be reduced to the former one by combining three sequent two-fold
nodes into one effective four-fold node at all the levels. In this case the
cubic domain of volume $l_{n}^{3}$ falls per each artery of level $n$. Let us
now consider individually three characteristic forms of the domain
$Q_{\mathcal{L}}$, a ball or a cube of size $\mathcal{L}$ ($d=3$), an
infinitely long cylinder of radius $\mathcal{L}$ ($d=2$), and a plane layer of
thickness $\mathcal{L}$ ($d=1$). For the ball or the cube, i.e. a region
bounded in three dimensions the probability that an artery of level $n$ passes
through the domain $Q_{\mathcal{L}}$ is about
\begin{equation*}
P_{l}^{(3e)}\sim \left( \frac{\mathcal{L}}{l_{n}}\right) ^{2}\,.
\end{equation*}
For the infinitely long cylinder
\begin{equation*}
P_{l}^{(2e)}\sim \left( \frac{\mathcal{L}}{l_{n}}\right)
\end{equation*}
and for the plane layer $P_{l}^{(1e)}\sim 1$. Multiplying $g_{l}(\mathbf{r})$
by the corresponding values of $P_{l}^{(e)}$ we get the result of averaging
the walker transition rate $g_{l}(\mathbf{r})$ over the possible realizations
of the vascular network embedding. The obtained result is written as
\begin{equation}
\left\langle g_{l}(\mathbf{r})\right\rangle \sim
\frac{D}{L_{n}}\frac{\mathcal{L}^{d}}{l_{n}^{2+d}}(T-T_{a}),  \label{90}
\end{equation}
where the value $d$ actually plays the role of the dimension of the space
inside which the temperature field can be considered. At the next step we
should sum the terms~(\ref{90}) over all the levels of the large artery tree.
However due to the strong increase of the terms~(\ref{90}), $\left\langle
g_{l}(\mathbf{r})\right\rangle \propto 2^{n(2+d)}$, as the level number $n$
increases the arteries of length $l\sim r$ mainly contribute to the value of
$\left\langle g_{l}(\mathbf{r})\right\rangle $. So the term describing the
fast heat transport with blood flow through large arteries from the domain
$Q_{\mathcal{L}}$ (located near the origin of the coordinate system) can be
written as
\begin{equation}
g(\mathbf{r})\sim \frac{D}{L_{n}}\int_{\mathcal{M}}d\mathbf{r}^{\prime }
\frac{T(\mathbf{r}^{\prime })}{\left| \mathbf{r}-\mathbf{r}^{\prime }\right|
^{5}}\,,  \label{opla}
\end{equation}
where $\mathcal{M}$ is the region containing the peripheral vascular network
as a whole and the integration in the three dimensional space over the
domain $Q_{\mathcal{L}}$ allows for all its three considered types.

Expression (\ref{opla}) together with the mean-filed bioheat
equation~(\ref{eq:BH}) enables us to write the following equation governing
the anomalous heat transfer in living tissue
\begin{equation}
\frac{\partial T}{\partial t}=D\nabla ^{2}T+\frac{D}{L_{n}}\int_{\mathcal{M}
}d\mathbf{r}^{\prime }\frac{T(\mathbf{r}^{\prime })-T(\mathbf{r})}{\left|
\mathbf{r}-\mathbf{r}^{\prime }\right| ^{5}+\ell _{v}^{5}}
-fj(T-T_{a})+q_{T}\,,  \label{eq:Final}
\end{equation}
where we have added directly the value $\ell _{v}$ in order to cut off the
spatial scales smaller than the length $\ell _{v}$ and ignored the difference
between the effective temperature diffusivity and the true one of the cellular
tissue. Equation~(\ref{eq:Final}) is the desired governing equation of the
anomalous fast heat diffusion in living tissue for the averaged temperature
field. It should be noted that the second term on the right-hand side of
equation~(\ref{eq:Final}) depends weakly on the blood perfusion rate.
Therefore, for the nonuniform distribution of the blood perfusion rate
$j(\mathbf{r},t)$ equation~(\ref{eq:Final}) holds also.

\subsection*{Anomalous temperature distribution under local strong heating}

Hyperthermia treatment as well as thermotherapy of small tumors of size about
or less 1~cm is related to local strong heating of living tissue up to
temperatures about 45$~^{\circ }$C or higher values. In this case the tissue
region heated directly, for example, by laser irradiation is also of a similar
size. Due to the tissue response to such strong heating the blood perfusion
rate can grow tenfold locally whereas in the neighboring regions the blood
perfusion rate remains practically unchanged \cite{Song84}. The feasibility of
such nonuniform distribution of the blood perfusion rate may be explained
applying to the cooperative mechanism of self-regulation in hierarchically
organized active media \cite{BOOK,SIAM}. Therefore in the region affected
directly the blood perfusion rate $j_{\text{max}}$ can exceed the blood
perfusion rate $j_{0}$ in the surrounding tissue substantially. In this case
the characteristic length of heat diffusion into the surrounding tissue is
about
\begin{equation}
\ell _{T}^{\ast }\sim \sqrt{\frac{D}{fj_{\text{max}}}}\,,  \label{add10}
\end{equation}
giving us also the minimal size $\mathcal{L}_{\text{mim}}$ of the region
wherein the tissue temperature increase $(T_{\text{max}}-T_{a})$ is mainly
localized. In the neighboring tissue the blood perfusion rate keeps up a
sufficiently low value $j_{0}$, which makes the heat propagation with blood
flow through large arteries considerable. Indeed, let us estimate the
temperature increase caused by this effect using the obtained
equation~(\ref{eq:Final}). The temperature increase $T(r)-T_{a}$ at a point
spaced at the distance $r>\mathcal{L}$ from the region (of size $\mathcal{L}$)
affected directly, i.e. the tail of the temperature distribution is mainly
determined by the anomalous heat diffusion and, so, is estimated by the
expression
\begin{equation}
T(r)-T_{a}\sim \frac{j_{\text{max}}}{L_{n}j_{0}}\frac{(\ell _{T}^{\ast })^{2}
\mathcal{L}^{3}}{r^{5}}\left( T_{\text{max}}-T_{a}\right) \,.  \label{haha}
\end{equation}
As seen from (\ref{haha}) for a sufficiently local and strong heating of the
tissue, i.e. when $\mathcal{L}\sim \ell _{T}^{\ast }$ and $j_{\text{max}}\gg
j_{0}$ the temperature increase at not too distant points such that $r\gtrsim
$ $\mathcal{L}$ can be considerable. Otherwise the anomalous heat diffusion is
ignorable.

\begin{acknowledgments}
This work was supported by STCU grant \#1675 and INTAS grant \#00-0847.
\end{acknowledgments}

\end{document}